\documentclass{emulateapj}

\shorttitle{Detection of PSR~B1509$-$58 and its pulsar wind nebula in MSH~15$-$52 using {\it Fermi}~-~LAT}
\shortauthors{Abdo et al.}

\begin{document}
\title{Detection of the energetic pulsar PSR~B1509$-$58 and its pulsar wind nebula in MSH~15$-$52 using the {\it Fermi}~-~Large Area Telescope}

\author{
A.~A.~Abdo\altaffilmark{2,3}, 
M.~Ackermann\altaffilmark{4}, 
M.~Ajello\altaffilmark{4}, 
A.~Allafort\altaffilmark{4}, 
K.~Asano\altaffilmark{5}, 
L.~Baldini\altaffilmark{6}, 
J.~Ballet\altaffilmark{7}, 
G.~Barbiellini\altaffilmark{8,9}, 
M.~G.~Baring\altaffilmark{10}, 
D.~Bastieri\altaffilmark{11,12}
% ...
}
% ...
\altaffiltext{2}{Space Science Division, Naval Research Laboratory, Washington, DC 20375, USA}
\altaffiltext{3}{National Research Council Research Associate, National Academy of Sciences, Washington, DC 20001, USA}
\altaffiltext{4}{W. W. Hansen Experimental Physics Laboratory, Kavli Institute for Particle Astrophysics and Cosmology, Department of Physics and SLAC National Accelerator Laboratory, Stanford University, Stanford, CA 94305, USA}
\altaffiltext{5}{Interactive Research Center of Science, Tokyo Institute of Technology, Meguro City, Tokyo 152-8551, Japan}
\altaffiltext{6}{Istituto Nazionale di Fisica Nucleare, Sezione di Pisa, I-56127 Pisa, Italy}
\altaffiltext{7}{Laboratoire AIM, CEA-IRFU/CNRS/Universit\'e Paris Diderot, Service d'Astrophysique, CEA Saclay, 91191 Gif sur Yvette, France}
\altaffiltext{8}{Istituto Nazionale di Fisica Nucleare, Sezione di Trieste, I-34127 Trieste, Italy}
\altaffiltext{9}{Dipartimento di Fisica, Universit\`a di Trieste, I-34127 Trieste, Italy}
\altaffiltext{10}{Rice University, Department of Physics and Astronomy, MS-108, P. O. Box 1892, Houston, TX 77251, USA}
\altaffiltext{11}{Istituto Nazionale di Fisica Nucleare, Sezione di Padova, I-35131 Padova, Italy}
\altaffiltext{12}{Dipartimento di Fisica ``G. Galilei", Universit\`a di Padova, I-35131 Padova, Italy}
% ...

\begin{abstract}
We report the detection of high energy $\gamma$-ray emission from the young and 
energetic pulsar PSR~B1509$-$58 and its pulsar wind nebula (PWN) in the composite 
supernova remnant SNR~G320.4$-$1.2 (aka MSH~15$-$52). 
Using 1 year of survey data with the {\it Fermi}-Large Area Telescope
(LAT), we detected pulsations from PSR~B1509$-$58 up to 1 GeV and
extended $\gamma$-ray emission above 1 GeV spatially coincident with
the PWN.  The pulsar light curve presents two peaks offset from the
radio peak by phases 0.96 $\pm$ 0.01 and 0.33 $\pm$ 0.02. New
constraining upper
limits on the pulsar emission are derived below 1 GeV and confirm a severe
spectral break at a few tens of MeV.  The nebular spectrum in the 1 --
100 GeV energy range is well described by a power-law with a spectral
index of (1.57 $\pm$ 0.17 $\pm$ 0.13) and a flux above 1 GeV of (2.91
$\pm$ 0.79 $\pm$ 1.35) $\times 10^{-9}$~cm$^{-2}$~s$^{-1}$.  The first
errors represent the statistical errors on the fit parameters, while
the second ones are the systematic uncertainties. The LAT spectrum of
the nebula connects nicely with Cherenkov observations, and indicates
a spectral break between GeV and TeV energies. 
\end{abstract}


\section{Introduction}
Pulsars and pulsar wind nebulae (PWNe) are believed 
to be sources of cosmic-ray electrons \citep[e.g.,][]{ken84,gae06}. 
Although hadronic $\gamma$-ray emission from TeV emitting PWNe 
has been suggested by many authors \citep[e.g.,][]{bed03, hor06}, 
most evidence indicates that $\gamma$-rays are generated 
via inverse Compton scattering of electrons accelerated in pulsar magnetospheres 
and at pulsar wind termination shocks.

The composite supernova remnant (SNR) G320.4$-$1.2 \citep[aka
  MSH~15$-$52;][]{cas81} is usually associated with the
rotation-powered radio pulsar PSR~B1509$-$58 (aka PSR~J1513$-$5908).
The 150~ms rotation period was discovered by the {\it Einstein}
satellite \citep{sew82} and soon thereafter confirmed in the radio
domain \citep{man82}.  With a large period derivative ($1.5 \times 10^{-12}\, \rm{s}\,\rm{s}^{-1}$), this
pulsar is one of the youngest and most energetic pulsars known
in the Galaxy with a characteristic age of 1700 years and a spin-down
power $\dot{E}$ of $1.8 \times 10^{37}$ erg~s$^{-1}$. The
inferred surface magnetic field is $1.5 \times 10^{13}$~G derived under the
assumption of a dipolar magnetic field.  The measurement of the pulsar
braking index $n = 2.839$ shows that this assumption is reasonable
\citep[e.g.,][]{liv05}. Therefore, the high magnetic field is not much
below the quantum-critical magnetic field of $4.413 \times
10^{13}$\,G, the domain of the so-called high-B-field pulsars and
magnetars.  The distance is estimated at $5.2\pm 1.4$~kpc using
HI absorption measurements~\citep{gae99}.  This is consistent
with the value of $4.2\pm 0.6$~kpc derived from the dispersion measure \citep{cor02}.  
PSR~B1509$-$58 has been studied
by all major X-ray and $\gamma$-ray observatories yielding a
broad-band spectral energy distribution and pulse profiles as a
function of energy.  Its detection by COMPTEL \citep[0.75~--~30~MeV,][]{kui99}
and non-detection with the Energetic Gamma-Ray Experiment Telescope (EGRET) in the 30~MeV~--~30~GeV
energy range, both aboard the Compton Gamma Ray Observatory ({\it CGRO}),
indicate an abrupt spectral break
between 10 and 30 MeV. This break is well below the break energies of
most $\gamma$-ray pulsars detected by {\it Fermi} which are typically
around a few GeV \citep{abd10a}. More recently, the detection of
pulsed $\gamma$-rays from PSR~B1509$-$58 at 4~$\sigma$ level above
100~MeV was reported by {\it AGILE} \citep{pel09}.

{\it Einstein} X-ray observations of MSH 15$-$52 revealed an elongated
non-thermal source centered on the pulsar~\citep{sew82}, later
confirmed by {\it ROSAT} and interpreted as a pulsar wind nebula
powered by PSR~B1509$-$58~\citep{tru96}.  This PWN, composed of
arcs and bipolar jets, is especially bright and extended in X-rays \citep{tam96,
gae02, for06, yat09}, and at very high energies \citep{sak00, aha05,
nak08}. The dimensions of the PWN as observed by ROSAT \citep{tru96} and H.E.S.S.
\citep{aha05} are 10' $\times$ 6' and 6.4' $\times$ 2.3' respectively.
The multi-wavelength emission of the PWN in MSH 15$-$52 can be
accounted for by synchrotron radiation from electrons within the PWN
and inverse Compton scattering on soft photons such as the cosmic
microwave background (CMB), the infrared (IR) and the optical interstellar
radiation field~\citep{aha05, nak08}.  This leptonic model requires a
broken power-law spectrum for the electrons.  However, large
uncertainties on the break energy remain due to the lack of
observations at corresponding wavelengths, namely IR or optical for
the synchrotron radiation and GeV $\gamma$-rays for the inverse
Compton component.  
The observations performed now with {\it Fermi} help constrain
the electron spectrum, in particular the break energy,
of the PWN of MSH~15$-$52 and provide new elements to the discussion
on the energetics.
 
Successfully launched in June 2008, the Large Area Telescope (LAT)
onboard the {\it Fermi} Gamma-ray Space Telescope covers the 20 MeV --
300 GeV energy range.  With its improved performance compared to its
predecessor EGRET, it offers the opportunity to search for high energy
pulsations of PSR~B1509$-$58, and, measuring the spectrum of the PWN
in MSH~15$-$52, to better constrain the emission models in pulsar
winds.  The results of 1-year observations of the pulsar
PSR~B1509$-$58 and its PWN are reported in the following sections.

\section{Radio Timing observations}
\label{radio}
With its large spin-down power, the pulsar PSR~B1509$-$58 is a good
candidate for $\gamma$-ray detection and is monitored by the LAT
pulsar timing campaign \citep{smi08} coordinated among \emph{Fermi},
radio and X-ray telescopes.

The ephemeris of the pulsar PSR~B1509$-$58 used in the analysis of the
\emph{Fermi}-LAT data is obtained using observations at 1.4 GHz made
with the 64-m Parkes radio telescope \citep{man08,wel09}.  A total of 42
time of arrivals (TOAs) were recorded between 2007 April 30 and 2009 August
29.  Radio observations simultaneous with the $\gamma$-ray data
make it possible to correct for the large drift in phase
caused by timing noise when constructing the $\gamma$-ray light curves.

The TEMPO2 timing package \citep{hob06} is then used to build the
timing solution from the 42 TOAs. We fit the TOAs to the pulsar
rotation frequency and its first two derivatives.
The fit further includes 3
harmonically-related sinusoids, using the ``FITWAVES'' option in the
TEMPO2 package, to flatten the timing noise. We used the value of DM~=~(252.5~$\pm$~0.3) cm$^{-3}$~pc
for the dispersion measure \citep{hob04}, given in the ATNF Pulsar Catalogue \footnote{http://www.atnf.csiro.au/research/pulsar/psrcat/}. 
The post-fit rms is 875 $\mu$s, or 0.6\% of the pulsar phase.  This timing
solution is used in the temporal analysis described in detail in
Section~\ref{phasos}.
 
\section{LAT description and data selection}
\label{lat}
The LAT is a high energy photon telescope sensitive
to $\gamma$-rays with energies from below 20 MeV to more than 300 GeV,
that detects photons through pair conversion. The photon incident direction
is derived by tracking the electron-positron pair in a high-resolution converter tracker
and the energy is measured with a CsI(Tl) crystal calorimeter. An 
anticoincidence detector identifies the background of charged particles \citep{atw09}.
In comparison to its predecessor EGRET, the LAT has a larger effective
area ($\sim$ 8000 cm$^{2}$ on-axis), a broader field of view ($\sim$
2.4 sr) and a superior angular resolution ($\sim$ 0.6$^{\circ}$ 68\%
containment at 1 GeV for events converting in the front section of the
tracker).

The analyses reported here are performed on 374 days of data taken in
survey mode (2008 August 04 -- 2009 August 13).  Events from the
``Diffuse'' class are selected, i.e. the highest quality photon data,
having the most stringent background rejection \citep{atw09}. In addition, we
exclude events with zenith angles greater than 105$^{\circ}$ 
to avoid contamination by $\gamma$-rays produced by cosmic-ray interactions in 
the Earth's atmosphere
and periods corresponding to a
rocking angle (i.e. the angle between the viewing direction of the LAT
and the zenith) larger than 43$^{\circ}$.

\section{Results}
\label{results}
\subsection{Gamma-ray emission from PSR~B1509$-$58}
\subsubsection{Light curves}\label{phasos}

We selected photons with an angle $\displaystyle \theta<
\max(5.12^{\circ} \times (\frac{E}{100 {\rm
    MeV}})^{-0.8},0.2^{\circ})$, where $E$ is the energy of
the photon, from the radio pulsar position, R.A.~$=(228.48175 \pm 0.00038)^{\circ}$,
Dec.~$=(-59.13583 \pm 0.00028)^{\circ}$ in J2000 \citep{kas94}
The energy-dependence of the integration radius is a satisfactory
approximation of the shape of the LAT Point Spread Function (PSF),
especially at low energies.

\begin{figure*}[ht!]
\begin{center}
\includegraphics[angle=0,scale=.8]{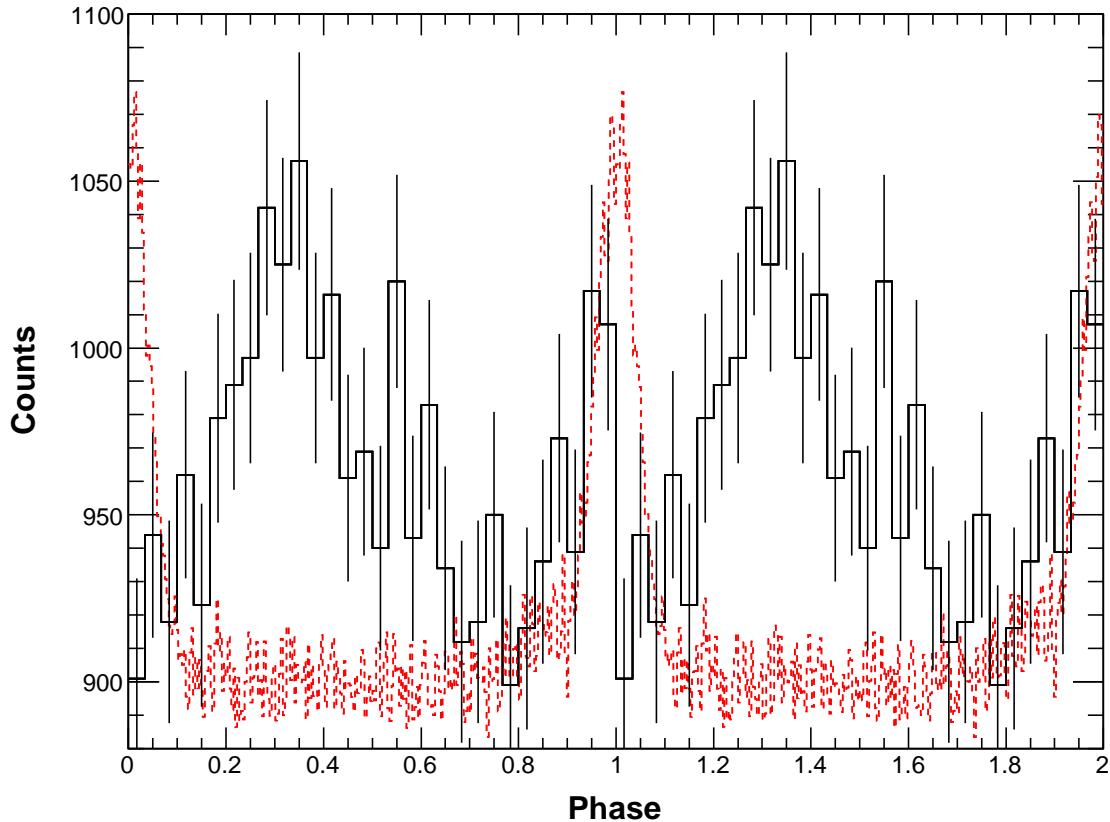}
\caption{\label{phaso_general} Light curve of the pulsar
  PSR~B1509$-$58 above 30 MeV within an energy-dependent circular
  region, as described in Section \ref{phasos}.  The light curve
  profile is binned to 1/30 of pulsar phase.  The radio profile (red
  dashed line) is overlaid in arbitrary units.  The main peak of the
  radio pulse seen at 1.4 GHz is at phase 0. Two cycles are
  shown.}
\end{center}  
\end{figure*}

Photons in this energy-dependent region are then phase-folded
using the timing solution described in Section \ref{radio}.  The
resulting $\gamma$-ray light curve for energies higher than 30
MeV, is presented in Figure~\ref{phaso_general}.  We have a total of
28966 $\gamma$-rays in the circular region of energy-dependent radius,
among which are 1267 $\pm$ 515 pulsed photons after background
subtraction. The radio profile (red dashed line) obtained from
the 42 radio observations included in the timing solution used for 
our analysis is overlaid in
Figure~\ref{phaso_general} for comparison. In this analysis, phase 0
is defined as the maximum of the main radio peak observed at 1.4 GHz.

\begin{figure*}
\begin{center}
\includegraphics[angle=0,scale=.7]{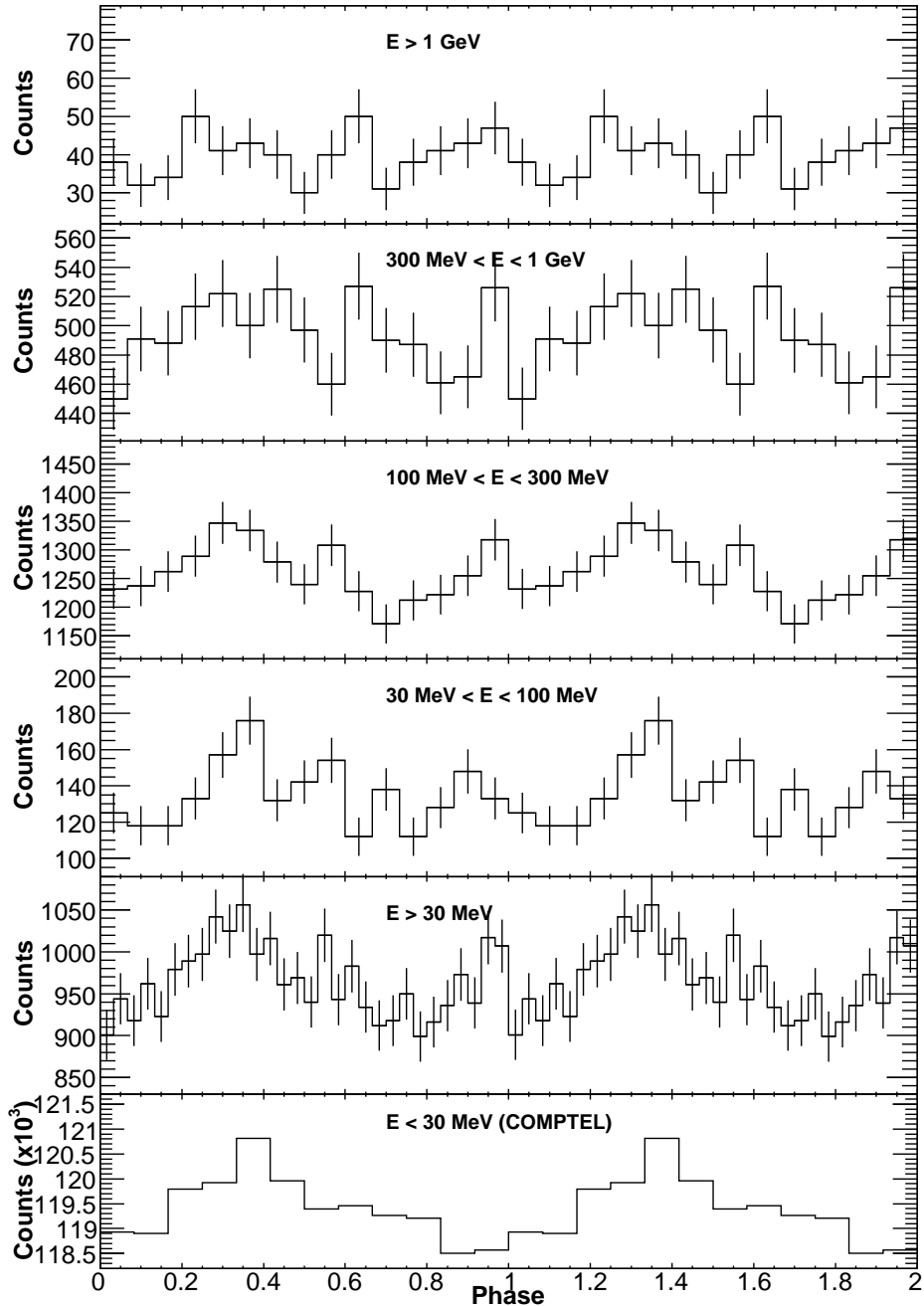}
\caption{\label{phaso_energy} Light curves of the pulsar
  PSR~B1509$-$58 in different energy bands within a circular region of
  energy-dependent radius.  From bottom to the top: COMPTEL
  \citep[0.75 -- 30 MeV;][]{kui99} and LAT profiles in 30 MeV--300
  GeV, 30 MeV--100 MeV, 100 MeV--300 MeV, 300 MeV--1 GeV, 1 GeV--300
  GeV energy bands are presented. Two cycles are
  shown. }
\end{center}
\end{figure*}

In Figure~\ref{phaso_general}, two peaks P1 and P2 can be observed at
phases 0.96 $\pm$ 0.01 and 0.33 $\pm$ 0.02 respectively.  The
uncertainty in the extrapolation of the radio pulse arrival time to
infinite frequency, as defined in \cite{Manchester and Taylor 1977},
is 0.004 in phase and can be neglected.  Hence, the peaks are
separated by $\Delta\phi$~=~0.37~$\pm$~0.02.  P1 and P2 are symmetric
and can be well modeled by Lorentzian functions of half-widths of
0.22~$\pm$~0.11 and 0.05~$\pm$~0.03 respectively.  We also notice that
the first $\gamma$-ray peak leads the radio main pulse by phase
0.04~$\pm$~0.01, as shown in Figure~\ref{phaso_general}, which
corresponds to a delay of (6~$\pm$~2)~ms.

Table~\ref{Htest} presents the value of the H-test as defined in
\cite{deJ89} and obtained in the 30 MeV~--~100 GeV, 30 MeV~--~100 MeV, 100
MeV~--~300 MeV, 300 MeV~--~1 GeV and 1 GeV~--~100 GeV energy bands using the
energy-dependent circular radius defined above.  The corresponding
light curves are presented in Figure~\ref{phaso_energy} along with the
light curve measured by COMPTEL in the 0.75~--~30~MeV energy range
\citep{kui99}. Within the error bars, the peak positions remain stable with energy. From
Figure~\ref{phaso_energy} and Table~\ref{Htest}, we notice that no
significant pulsation can be detected above 1 GeV. Using 374 days of
data in survey mode, a H-test value of 31.34 is obtained in the 30 MeV~--~100 
GeV energy range, corresponding to a significance of
4.51~$\sigma$.

\begin{table}[h!]
\begin{center}
\begin{tabular}{lrc}
\hline
\hline
Energy band  & H-test & Significance \\
(GeV) &  & ($\sigma$) \\
\hline
0.03 -- 100	&  31.34 & 4.51	\\
0.03 -- 0.1 	&  15.42 & 3.07	\\
0.1 -- 0.3 	&  15.60 & 3.09	\\
0.3 -- 1.0 	&  4.66 & 1.42	\\
1.0 -- 100 	&  0.06	& 0.03	\\
\hline
\end{tabular}
\caption{\label{Htest} 
  Results of the periodicity test applied to PSR~B1509$-$58 
  using the energy-dependent region defined in Section~\ref{phasos}}
\end{center}
\end{table}

\subsubsection{Spectral analysis of PSR~B1509$-$58}\label{pulsarspectrum}
A spectral analysis of the pulsar is performed in the 100 MeV -- 1 GeV
energy range using a maximum-likelihood method \citep{mat96}
implemented in the \emph{Fermi} SSC science tools
as the {\it gtlike} code.  This tool fits a source model to the data
along with models for the instrumental, extragalactic and Galactic
backgrounds.  In the following spectral analysis, the Galactic diffuse
emission is modeled using the ring-hybrid model {\it
  gll\_iem\_v02.fit}.  The instrumental background and the
extragalactic radiation are described by a single isotropic component
with a spectral shape described by the tabulated model {\it
  isotropic\_iem\_v02.txt}.  These models and their detailed description are released by the LAT
Collaboration \footnote{$Fermi$ Science Support Center:
  http://fermi.gsfc.nasa.gov/ssc/}.  Sources near the pulsar
PSR~B1509$-$58 found above the background with a statistical
significance larger than $5 \, \sigma$ are extracted from
the source list given in \cite{abd10b}, and are taken into account in this study. We use
P6\_V3 post-launch instrument response functions that take into
account pile-up and accidental coincidence effects in the detector
subsystems \footnote{ See http://fermi.gsfc.nasa.gov/ssc/data/analysis/documentation/Cicerone/Cicerone\_LAT\_IRFs/IRF\_overview.html for more details}.

\begin{figure*}
\begin{center}
\includegraphics[angle=0, scale=.60]{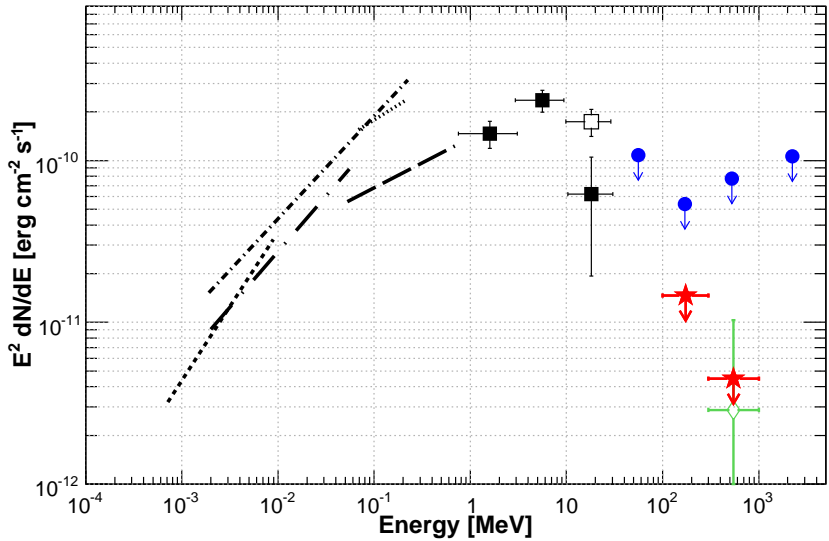}
\caption[]{Spectrum of PSR~B1509$-$58 from soft X-rays to $\gamma$-rays.
  2~$\sigma$ upper limits derived with {\it gtlike} from LAT observations are
  represented by red stars. 
  The green diamond represents the integrated flux derived from the light curves and 
  estimates of the effective area and exposure in the 300~MeV~--~1~GeV energy range, 
  as described in section \ref{pulsarspectrum}.
  The filled squares are the COMPTEL flux
  points derived from the excess counts in the 0.15 -- 0.65 phase
  range while the open square represents the 10 -- 30 MeV flux in the
  0.15 -- 0.65 phase interval above the spatially determined
  background.  The blue circles are the 2~$\sigma$ upper limits for
  the total fluxes obtained by EGRET.  The different lines represent
  the best spectral fit measured by {\it ASCA} (short-dashed line, 0.7-10 keV), {\it Ginga} (dot-long-dashed line, 2-60 keV), {\it CGRO}-OSSE (long-dashed line, 50-750 keV), {\it WELCOME} (dotted line, 94-240 keV)
  and {\it RXTE} (dot-short-dashed line, 2-250 keV) \citep[][and references
    therein]{kui99}.
    }\label{pulsar_spectrum}
\end{center}
\end{figure*}

Despite the detection of pulsations, no significant $\gamma$-ray
emission can be observed at the position of the pulsar using 374 days
of LAT data.  Indeed, most of the weak signal detected on
PSR~B1509$-$58 is observed at low energy (below 300~MeV) where the LAT
angular resolution is large in comparison to the distance that
separates our source of interest from the bright pulsar
PSR~J1509$-$5850 (less than 0.8$^{\circ}$).  This renders the spectral
analysis extremely complex.  Therefore, only 2~$\sigma$ upper
limits can be derived and are presented in
Figure~\ref{pulsar_spectrum}. 
As an attempt to evaluate the flux of PSR~B1509$-$58, pulsed excess 
counts were derived from the light curves (presented in 
Figure~\ref{phaso_energy}) in the 100~--~300~MeV and 300~MeV~--~1~GeV
energy bands, as well as the corresponding effective area.
The rapid increase of the effective area in the 100~--~300~MeV energy range \citep{atw09} makes
the analysis highly dependent of the assumed spectral shape and yields very large errors on the flux estimate.
Therefore, only the result obtained in the 300~MeV~--~1~GeV energy range \citep{atw09}, 
where the effective area is more stable, is represented in Figure~\ref{pulsar_spectrum}.
The upper limit derived with {\it gtlike} in this energy band is consistent with the 
integrated flux estimated from the number of pulsed photons.
The overall spectrum in the 1 keV -- 1 GeV energy
range indicates a very low cut-off or break energy in the pulsar
spectrum, as suggested by \cite{kui99}.  The implications of such
upper limits on the emission models in pulsar magnetospheres are
further discussed in Section~\ref{discussion_pulsar}.

\begin{table*}[!]
\begin{center}
\begin{tabular}{cccccc}
\hline
\hline
Spatial model  & Galactic & Galactic & Error & Radius & Test Statistics\\
             & longitude & latitude  & (degrees) & (degrees)& \\
             & (degrees) & (degrees) &          &      & \\
\hline
Point Source & 320.288 & -1.209 & 0.028 &                   & 44.7\\
Gaussian     & 320.275 & -1.266 & 0.051 & 0.146 $\pm$ 0.023 & 67.6\\
Uniform disk & 320.270 & -1.271 & 0.061 & 0.249 $\pm$ 0.047 & 69.4\\
\hline
\end{tabular}
\caption{Position of the centroid, extension and significance of the
  PWN in MSH~15$-$52 obtained with {\it Sourcelike} applied to the LAT
  data.}\label{Morpho}
\end{center}
\end{table*}

\subsection{High energy analysis of the PWN in MSH~15$-$52}
\subsubsection{Morphology}\label{counts_maps}
\begin{figure*}
\begin{center}
\includegraphics[angle=0, scale=.53]{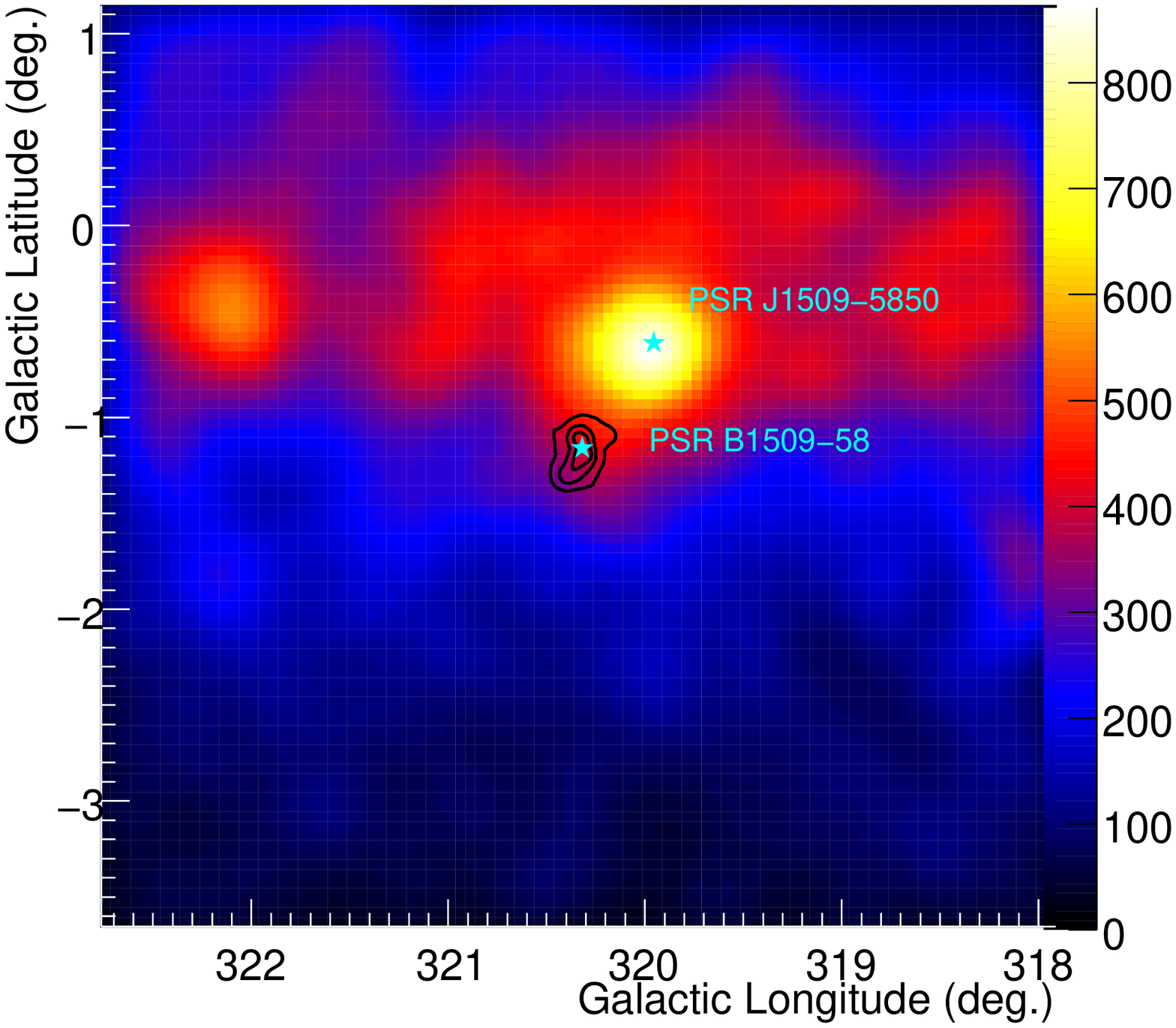}
\includegraphics[angle=0, scale=.53]{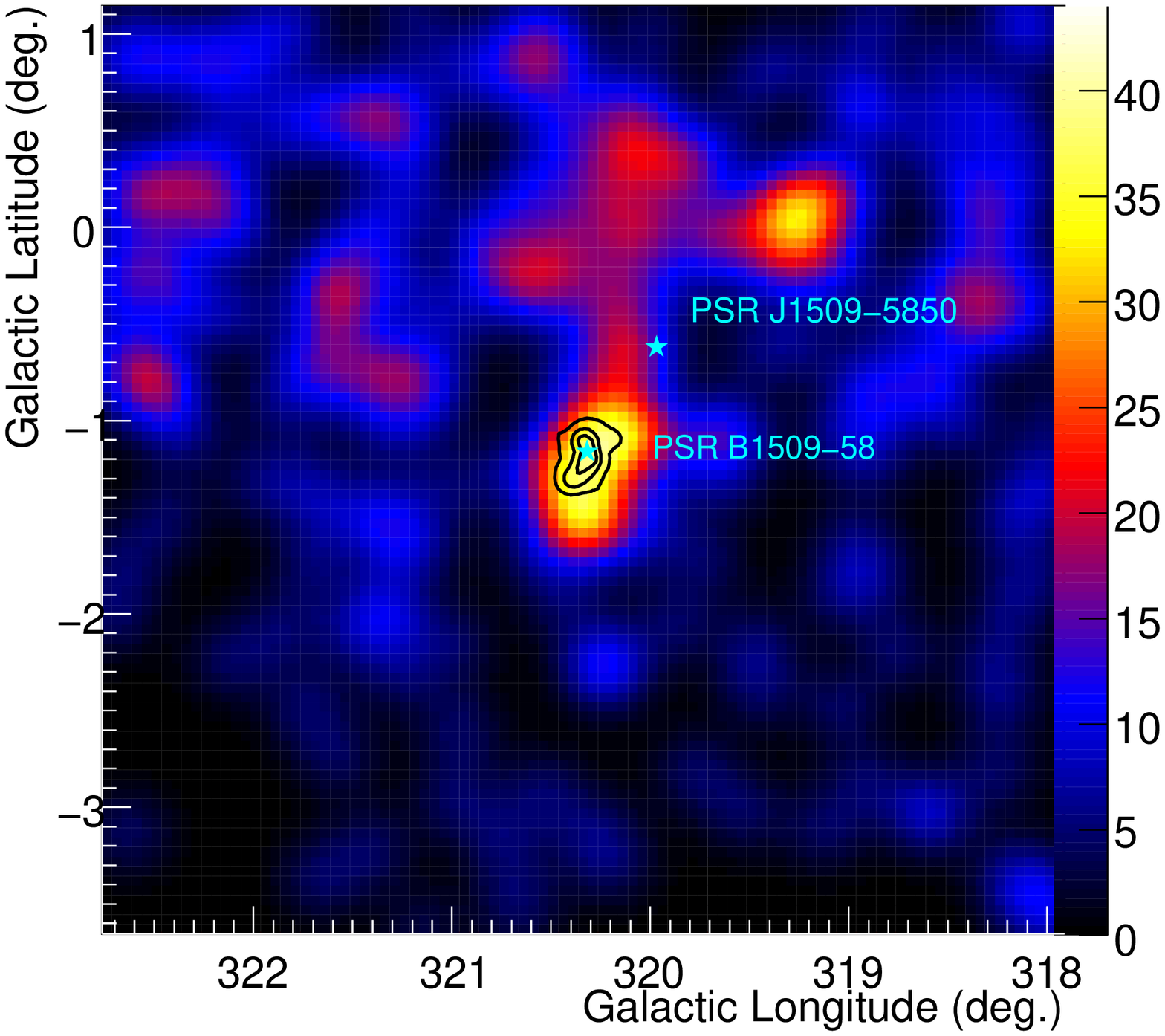}
\caption{\label{CMAP} Smoothed counts maps in arbitrary units
  of the region around MSH~15$-$52 above 1 GeV ({\it top}) and 10 GeV
  ({\it bottom}) in Galactic coordinates, binned in square pixels of
  side-length 0.05$^{\circ}$. The maps are smoothed with a
    Gaussian of $\sigma$~=~0.15$^{\circ}$. The H.E.S.S. \citep{aha05}
  contours of the PWN in MSH~15$-$52 are overlaid in black for
  comparison. The positions of the pulsars PSR~J1509$-$5850 and
  PSR~B1509$-$58 are marked by blue stars.}
\end{center}
\end{figure*}

Figure~\ref{CMAP} presents the smoothed counts maps of the region
around MSH~15$-$52 in Galactic coordinates above 1 GeV (top panel) and
10 GeV (bottom panel) and binned in square pixels of side-length
0.05$^{\circ}$.  The H.E.S.S. \citep{aha05} contours have been
overlaid in black for comparison.  At low energies, the emission is
essentially dominated by the bright nearby pulsar PSR~J1509$-$5850
\citep{abd10a}, marked by a blue star, whereas the significant
$\gamma$-ray emission above 10 GeV is spatially coincident with the
nebula in MSH~15$-$52.  The position of the pulsar PSR~B1509$-$58 is
also marked with a blue star.

An analysis tool, {\it Sourcelike}, developed by the LAT Collaboration
allows us to estimate the position and the size of the source,
assuming a spatial and spectral model for the diffuse emission and
different morphologies: a point source, a Gaussian shape and a uniform
disk. In this method, the likelihood is iterated to the data set
assuming spatial source models, taking into account nearby sources,
Galactic diffuse and isotropic components in the fits \citep{abd10c},
as described in Section~\ref{pulsarspectrum}.

Morphological studies are performed above 6.4 GeV, assuming the three
spatial hypotheses mentioned above. The choice of this energy is
motivated by the better angular resolution, the non-contamination from
the Galactic diffuse emission, and nearby bright sources such as the
pulsar PSR~J1509$-$5850. The positions, extensions and the
corresponding errors, as well as the Test Statistics (TS) obtained for each
hypothesis are summarized in Table~\ref{Morpho}. The TS values are
obtained as TS~=~2($L_1 - L_0$), where $L_0$ and $L_1$ are the
values of the log-likelihood obtained by null hypothesis and each
source hypothesis, respectively.  In view of the errors of
localization, the fit positions are compatible with each other, and
bests fits are obtained either with a uniform disk (TS of 69.4) or a
Gaussian distribution (TS of 67.6).

The differences in Test Statistics between the two extended shapes and
the point source hypotheses: TS$_{ext}$ = 24.7 and TS$_{ext}$ = 22.9
for the uniform disk of extension $\sigma$ $\sim$ 0.25$^{\circ}$ and
the Gaussian shape of extension $\sigma$ $\sim$ 0.15$^{\circ}$
respectively with relation to a point source, indicate a significant
extension for the LAT source spatially coincident with the PWN in
MSH~15$-$52. The conversion of the extensions obtained for the uniform
disk and Gaussian distribution into a rms value gives results
consistent with each other. In the following analyses, the spectral
results obtained for the extended scenarios are presented. For
comparison, the results derived assuming a point source hypothesis are
also quoted.

\subsubsection{Spectral analysis}\label{nebula}
The following spectral analyses are performed using {\it gtlike}. {\it
Sourcelike}, described in Section~\ref{counts_maps}, is also run as
a cross-check and gives compatible results. The models used to
describe the Galactic, extragalactic and instrumental components are
mentioned in Section~\ref{pulsarspectrum}. Sources near the PWN with
a statistical significance larger than $5 \, \sigma$ are extracted
from \cite{abd10b}, and are taken into account in this study.

Since the pulsar PSR~B1509$-$58 is seen only below 1~GeV, the
$\gamma$-ray photons in the 1 -- 300~GeV energy range, in a
20$^{\circ}$~$\times$~20$^{\circ}$ square centered on the the pulsar
radio position and coming from the entire pulse phase interval are
selected.

The spectral analysis is performed above 1 GeV assuming different
morphologies for the source: a point source, a Gaussian shape and a
uniform disk using the positions and extensions summarized in
Table~\ref{Morpho}. A spectral fit using the H.E.S.S. image of
  the PWN \citep{aha05} as a template is also performed.  In this
energy range, for all spatial distributions, the LAT spectrum can be
well modeled by a simple power-law:
\begin{equation}
  \frac{dN}{dE} \propto \Bigl(\frac{E}{1\rm GeV}\Bigr)^{-\Gamma} \, \rm{cm^{-2} \, s^{-1} MeV^{-1}} 
\end{equation}
where $\Gamma$ is the photon index of the spectrum.  The integrated
fluxes and spectral indices of the source obtained for different
spatial hypotheses are summarized in Table~\ref{spectral_results}.
The best fit is obtained for a uniform disk hypothesis, favored with
respect to the Gaussian, H.E.S.S. template and point source
morphologies with differences in TS of 3.9, 13.7 and 32.9 respectively.  The
LAT spectrum for a disk hypothesis is well described by a power-law
with a spectral index of (1.57 $\pm$ 0.17 $\pm$ 0.13) and a flux above
1 GeV of (2.91 $\pm$ 0.79 $\pm$ 1.35) $\times
10^{-9}$~cm$^2$~s$^{-1}$. The first errors represent the statistical
error on the fit parameters, while the second ones are the systematic
uncertainties.

Three different uncertainties can affect the LAT spectrum estimation,
as described in \cite{abd10d}.  The first one is due to the
uncertainty in the Galactic diffuse emission since MSH~15$-$52 is
located close to the Galactic plane.  Different versions of the
Galactic diffuse emission generated by GALPROP \citep{str04a,
    str04b} were used to estimate this error.  The difference with
the best diffuse model is found to be less than 6\%.  This implies
systematic errors on the fluxes of 33\% above 1 GeV.  The second
systematic is related to the morphology of the LAT source.  As
described in Table~\ref{Morpho}, the Gaussian, disk hypotheses and the
H.E.S.S. template match the gamma-ray morphology quite well.  The fact
that we can not decide which one is better adapted induces an
additional systematical error on the flux of the order of 24\% above 1
GeV.  The third systematic is produced by the uncertainties in the LAT
instrument response functions (IRFs). We bracket the energy-dependent
effective area with envelopes above and below the nominal curves by
linearly connecting differences of (10\%, 5\%, 20\%) at $\log(E/1~{\rm
  MeV})$ of (2, 2.75, 4) respectively, which yields additional errors
on the flux and spectral index.

\begin{figure*}
\begin{center}
\includegraphics[angle=0, scale=.60]{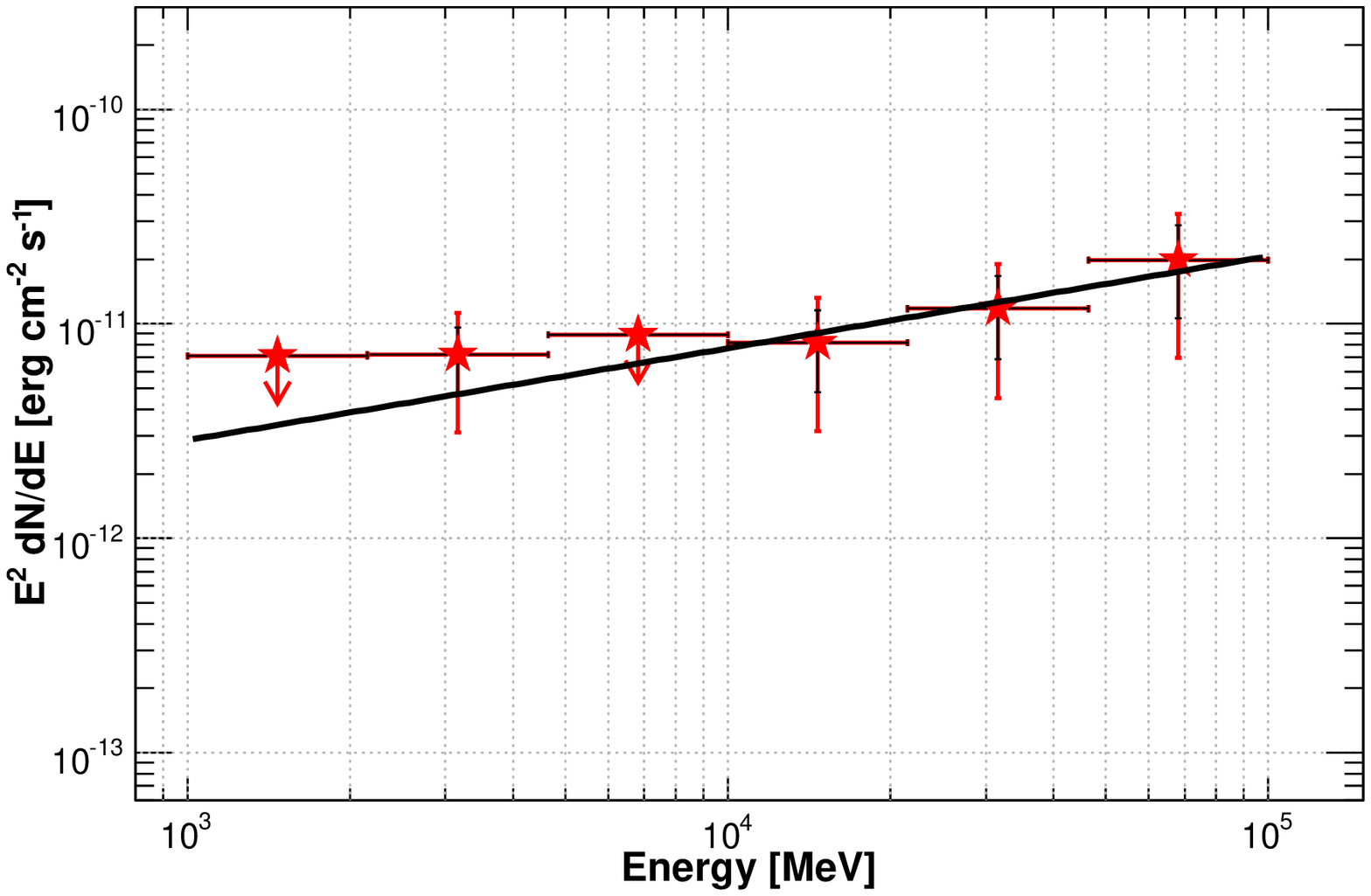}
\caption{\label{SED}
  Spectral energy distribution of the PWN above 1 GeV.
  The black line represents the results of the fit on the 1 GeV -- 100 GeV energy band.
  The spectral points are obtained using the model-independent maximum likelihood method
  described in Section~\ref{nebula}. 2~$\sigma$ upper limits are computed when the statistical 
  significance of the energy interval is lower than 3~$\sigma$.
  The statistical errors are shown in black, 
  while the red lines take into account both the statistical and systematic errors.
  Horizontal bars delimit the energy intervals. }
\end{center}
\end{figure*}

The spectral energy distribution (SED) of the PWN in the case of a
uniform disk hypothesis is presented in Figure~\ref{SED}.  The {\it 
Fermi}-LAT spectral points were obtained by dividing the 1~--~100~GeV
range into 6 logarithmically-spaced energy bins and performing
a maximum likelihood spectral analysis in each interval, assuming a
power-law shape for the source.  These points provide a
model-independent maximum likelihood spectrum, and are overlaid with
the fitted model over the total energy range (black line).  2~$\sigma$
upper limits are derived in energy bands where the significance
level of the signal is lower than 3~$\sigma$.

\begin{table}[ht!]
\begin{center}
\begin{tabular}{ccc}
\hline
\hline
Spatial model  & Flux above 1 GeV & Spectral index\\
               & (${\rm 10^{-9} \, cm^{-2} \, s^{-1}}$) & \\
\hline
Point Source 	& 2.00 $\pm$ 0.76 & 1.57 $\pm$ 0.24 \\
Gaussian 	& 3.01 $\pm$ 0.81 & 1.58 $\pm$ 0.17 \\
Uniform disk 	& 2.91 $\pm$ 0.79 & 1.57 $\pm$ 0.17 \\
H.E.S.S.        & 2.22 $\pm$ 0.77 & 1.52 $\pm$ 0.21 \\
\hline
\end{tabular}
\caption{Spectral parameters of the PWN obtained with {\it gtlike} for different spatial models.
  Statistical errors only are quoted.}\label{spectral_results}
\end{center}
\end{table}

\section{Discussion}

\subsection{High energy emission from PSR~B1509$-$58}\label{discussion_pulsar}
The improved performance of the \emph{Fermi}-LAT compared to its
predecessor EGRET allows the first detection of pulsations from
PSR~B1509$-$58 up to 1~GeV with a light curve presenting two peaks at
phases 0.96 $\pm$ 0.01 and 0.33 $\pm$ 0.02 as seen in Figure~\ref{phaso_general}.
The alignment of the broad peak (P2) at phase $\phi = 0.33$ is in general agreement with 
measurements of the phase of the main pulse by other high-energy experiments,
e.g. 0.38 $\pm$ 0.03 with COMPTEL \citep[10~--~30~MeV,][]{kui99}, 
0.32 $\pm$ 0.02 with {\it CGRO}-BATSE and {\it CGRO}-OSSE \citep[20~--~500~keV,][]{ulm93}
and $\sim$~0.35 with AGILE \citep[0.1~--~30~GeV,][]{pel09}. 
In X-rays, the peak shifts
towards phase 0.24~--~0.27 \citep[e.g.][]{kaw01, rot98, cus01}.
We do not detect the suggested narrow
pulsed component located at phase $\sim$~0.85 reported by \cite{kui99}
using combined COMPTEL and EGRET data in the 10~--~100~MeV energy range. However, we
observe a pulse component at phase 0.96 $\pm$ 0.01. Considering the marginal
significance of the COMPTEL/EGRET pulse, these peaks might be one single component.
AGILE also observed a
possible peak at phase $\sim$~0.85 \citep{pel09}, which appears to be
broader than the COMPTEL pulse. The detection of the narrow pulse only above 10 MeV
strongly suggests a harder spectrum for this component than for the broad one.

The {\it Fermi} pulse shapes can be modeled with symmetric Lorentzian
functions with half widths of 0.08 $\pm$ 0.06 and 0.21 $\pm$ 0.09
for P1 and P2, respectively. Below MeV energies, the broad component
is asymmetric. Its profile can be described with 2 Gaussians
components \citep[e.g.][]{kui99, cus01}. These components peak at
phases $\sim$~0.25 and $\sim$~0.39 with widths of 0.056 and 0.129,
respectively. Above a few tens of MeV, it seems that the first of
the two components is no longer contributing to the pulse profile,
causing the apparent 'shift', as observed by \citet{kui99}: the
COMPTEL broad peak can be described with one single
component. Within errors, the widths of the broad peak measured by
COMPTEL and {\it Fermi} are in agreement.

The pulse component observed at phase 0.96 $\pm$ 0.01 leads the 1.4
GHz radio pulse by 0.04 $\pm$ 0.01 in phase. This feature is quite
remarkable as so far all radio-loud pulsars excluding millisecond
pulsars detected by {\it Fermi}
present phase lags with respect to the radio pulse \citep{abd10a}. 
However, the first $\gamma$-ray peak of the Crab pulsar precedes
the main radio peak, but the high-energy peak lags the radio
precursor \citep[e.g.][]{abd10e}. In the case of PSR~B1509$-$58, a weak
component was reported by \citet{cra01} at 1351 MHz, which precedes the
main pulse by $\sim$~0.14 in phase. This might be a precursor as
is seen in the Crab, but this has to be confirmed before we draw
conclusions based on this possible feature.

Before {\it Fermi} launch, one of the major open questions dealt with
the zone of gamma-ray production in pulsar magnetospheres. 
There are two classes of models that
differ by the location of the emission region. First, polar-cap (PC)
models place the emission near the magnetic poles of the neutron star
\citep{Daugherty and Harding 1996}. The second class is formed by the
outer-gap (OG) models \citep[e.g.][]{che86, Romani 1996},
where the emission extends between the null charge surface and the
light cylinder, and by the two-pole caustic (TPC) models
\citep[e.g.][]{Dyks and Rudak 2003} which might be realized in slot
gap (SG) acceleration models \citep[e.g.][]{Muslimov and Harding
  2004}, where the emission takes place between the neutron star
surface and the light cylinder along the last open field line. After
one year of {\it Fermi} observations, the high-altitude models 
seem to be favored, even if these models do not work for
all pulsars. From previous high-energy observations, it was not
obvious beforehand whether the high-energy emission from PSR~B1509$-$58 could be
described by surface or outer-gap emission.

The $\gamma$-ray pulse profiles can be used to constrain the geometry
of the pulsar. \cite{wat09} simulated
a population of young spin-down-powered pulsars for vacuum-dipole
magnetospheres.
The peak separation of the $\gamma$-ray pulse profile presented in 
Figure~\ref{phaso_general} and the radio lag
can yield constraints on the viewing angle $\zeta$ and the magnetic
inclination $\alpha$. 

Using the heuristic law for the $\gamma$-ray luminosity \citep{wat09}:
\begin{equation}
L_{\gamma} \approx \eta \dot{E} \approx C \times \left(\frac{\dot{E}}{10^{33} \, \rm{erg \, s^{-1}}}\right)^{1/2} \times \rm{10^{33}  \, erg \, s^{-1}}
\end{equation}
with C a slowly varying function of the order of unity, the $\gamma$-ray efficiency $\eta$ is estimated at 0.007.
Assuming a $\gamma$-ray peak separation of 0.37 $\pm$ 0.02 and a $\gamma$-ray efficiency of 
$\eta = 0.01$, the closest value to the real $\gamma$-ray efficiency that can be found in \citet{wat09}, yields a tight constraint for 
$\zeta$ in the range $64^{\circ} - 70^{\circ}$ and $\alpha$ values of $45^{\circ} - 65^{\circ}$
in the framework of the OG model. The pulse profile can also be explained with the TPC model
assuming $\zeta$ of $50^{\circ} - 65^{\circ}$ and $\alpha$ of $45^{\circ} - 65^{\circ}$.
However, the radio delay with respect to the first $\gamma$-ray peak could not be explained neither
by the TPC nor the OG models. 

Constraints on $\alpha$ and $\zeta$ are also possible with
radio-polarization measurements \citep[e.g.][]{Radhakrishnan and
  Cooke 1969, lyn88}.  Polarization measurements by \cite{cra01}
show highly linearly polarized signals of 97\% and 94\% at 660 MHz
and 1351 MHz, respectively. However, the position angle shows only a
shallow swing, which suggests a large magnetic-pole impact angle
$\beta = \zeta - \alpha$. Magnetic inclination angles larger than
$60^{\circ}$ are excluded, in agreement with the maximum values of 
$\beta$ derived using the OG and TPC models of $25^{\circ}$ and $20^{\circ}$,
respectively. 

From the spectral analysis, the stringent upper limits of the pulsed spectrum measured by
{\it Fermi}-LAT and presented in Figure~\ref{pulsar_spectrum} 
confirm the spectral break between 10 and 30 MeV suggested by \cite{kui99}.
However, the new 2~$\sigma$ upper limits obtained with {\it Fermi} give
more stringent constraints than EGRET's one. The emission models can now be
  tested against these constraints.

\citet{Romani 1996} modeled the high-energy emission based on
curvature radiation-reaction-limited charges in the outer
magnetosphere and argued that, for high-field pulsars such as
PSR~B1509$-$58, synchrotron flux will dominate the 
emission in the 100~keV~--~10~MeV band, and more
specifically the GeV band curvature component.  The reason for this is
that high altitude two-photon pair creation in collisions between X-rays
and hard gamma-rays is prolific, enhanced by aberration effects. This
process will curtail super-GeV flux while permitting synchrotron
emission at lower energies by electrons with $\sim$ 0.3~--~10~GeV energies.
In this way, the radiative power is transferred in cascading from the
super-GeV band to a $\lesssim 10$MeV synchrotron window, i.e. generating
a spectrum peaking below the {\it Fermi}-LAT energy range.
Based on the three dimensional
outer-magnetosphere model of pulsars proposed by~\cite{cheng},
\cite{zhang} calculated the light curves and spectra of PSR~B1509$-$58
assuming $\alpha = 65^{\circ}$ and $\zeta = 75^{\circ}$.
Their light
curve presents a single broad peak comparable with the RXTE pulse
profile, though a bit too narrow \citep{rot98}. The resulting spectrum is characterised by a simple power-law from the
soft X-ray band to a few hundred keV, with a photon index of 1.5 and a cut-off below 1~MeV. This overall shape agrees well with the multi-wavelength data available even though the proposed cut-off lies at an energy slightly smaller than that proposed by~\cite{kui99} which is consistent with the new LAT data. Furthermore, their light curve presents a single broad peak while a second peak is observed with the LAT data above 30~MeV.

\cite{harding} argued that the polar cap model could explain the
low spectral cut-off observed by CGRO for PSR~B1509$-$58. Curvature
emission at low altitudes would naturally appear at energies above
10 GeV, which was obviously not seen by EGRET and cannot be
discerned in the LAT data presented here.  Spectral attenuation by
magnetic pair creation $\gamma B\to e^+e^-$ would be extremely
effective at energies above 100 MeV in the scenario of
\cite{harding}.  Using Eq.~[1] of \cite{bar04}, the maximum
photon energy consistent with magnetic pair production transparency
at altitude $r$ along the last open field line is $E_{\rm max}\sim
1.76 (B_{12})^{-1}\, P^{1/2}\, (r/R_{\ast})^{7/2}$~GeV, for a
surface polar field strength of $B_0=10^{12}B_{12}$G and stellar
radius $R_{\ast}\sim 10^6$cm.  $B_0\sim 1.5 \times 10^{13}$~G and
$P=0.15$\,s then set $E_{\rm max}\sim 45 (r/R_{\ast})^{7/2}$~MeV
for the energy of the $\gamma B$ pair creation turnover. Even at
the stellar surface, this estimate is too high to accommodate the
EGRET upper limits and COMPTEL data downturn at around 10 MeV as seen in
Figure~\ref{pulsar_spectrum}.

However, as emphasized by \cite{harding}, magnetic photon splitting, a
quantum-electrodynamic process important only for magnetic fields
approaching the quantum critical value $B_{cr} = 4.413 \times
10^{13}$~G, can attenuate $\gamma$-rays emitted near the surface of
strongly magnetized pulsars. \cite{harding} showed that photon
splitting will be important for $\gamma$-ray pulsars having a surface
magnetic field larger than $0.3 B_{cr}$, where the splitting
attenuation lengths and escape energies become comparable to or less
than those for pair production.  Specifically, they demonstrated that
attenuation due photon splitting would reduce $E_{\rm max}$ to nicely
accommodate the CGRO observations, but only if the emission was
predominantly at $r\sim R_{\ast}$ and its co-latitude was consistent
with a standard polar cap in PSR~B1509-58. The new {\it Fermi}-LAT
data with the combined soft X-ray to soft $\gamma$-ray (COMPTEL)
spectral points confirm this picture: the polar-cap model is
spectroscopically viable for this pulsar, but subject to the strong
constraint of emission at the magnetic co-latitude of the rim,
i.e. $\sim 2^{\circ}$ as proposed by \cite{kui99}.  Higher altitudes 
and accompanying larger co-latitudes will push $E_{\rm max}$ to energies above 50 MeV.
Accordingly, the LAT suggestion of modest pulsations up to energies
almost as high as 1 GeV indicates that some portion of this emission
might emanate from altitudes well above the stellar surface, where 
photon splitting will play a minimal role.

Although there are only upper limits on the pulsed spectrum in the LAT 
energy range at this point, the emission from both components of the 
light curve seems to extend to 1 GeV. The broad peak at phase 0.33 is 
consistent with outer magnetosphere geometry and is also roughly in phase
with the COMPTEL peak. If it is assumed that the radio peak arises at small magnetic
colatitudes, then the radio/soft $\gamma$-ray phase separation suggests that both 
the {\it Fermi} and COMPTEL emission components at this phase originate in the outer magnetosphere, 
but given the sharp cut-off just above COMPTEL energies they must have 
different mechanisms. The narrow {\it Fermi} peak at phase 0.96 just leading
the radio peak has no counterpart at COMPTEL energies but given its extension
to 1 GeV and the magnetic pair and photon splitting attenuation limits
discussed above, this peak must also originate in the outer magnetosphere.
However, its phase location is not easily explained by current outer 
magnetosphere gap models.  High altitude pair-starved polar cap emission 
has been shown to produce a single peak just leading the radio pulse 
\citep{ven09} and could potentially explain this
component of the light curve at a similar $\alpha$ and $\zeta$ range.  
However, explaining both {\it Fermi} light curve peaks would require both 
pair-starved and non-pair-starved (gap) models to co-exist.

One might expect the high magnetic field of PSR~B1509$-$58 to be
the main reason for its unique behavior. This seems not to be the
case if one compares PSR~B1509$-$58 with the other pulsar Fermi has
detected with an inferred surface magnetic field above $10^{13}$~G :
the nearby (1.4 kpc) PSR~J0007$+$7303 \citep{abd08, abd10a} has a 
period of 314 ms and a period derivative of
$3.61 \times 10^{13}$ s\,s$^{-1}$. Its inferred surface magnetic is
$1.1 \times 10^{13}$~G. The similar magnetic fields of PSR~J0007$+$7303
and of PSR~B1509$-$58 seem not be the criterion for similar
behavior as they appear to be very different. Indeed, PSR~J0007$+$7303
belongs to the top ten brightest $\gamma$-ray pulsars and has a hard
power-law spectrum with photon index $\Gamma = 1.38$ with a spectral
cutoff energy at 4.6 GeV. Its spin-down power ($4.5 \times
10^{35}$ erg\,s$^{-1}$) is almost two orders of magnitude less than
PSR~B1509$-$58. Also PSR~J0007$+$7303 shows no strong pulsed radio and
X-ray emission like PSR~B1509$-$58 \citep{hal04}, is about ten times older
and has a higher $\gamma$-ray efficiency than PSR~B1509$-$58. Finally, the
pulse profiles and spectral behavior of PSR~J0007$+$5303 can be explained
nicely with an outer-magnetosphere model unlike PSR~B1509$-$58.
\subsection{Constraints on the emission models in the nebula}
\label{discussion:neb}
High energy photons coming from pulsar magnetospheres are usually
expected to have a power-law spectrum with an exponential cut-off at a
few GeV \citep{abd10a}. \cite{kui99} suggested that
PSR~B1509$-$58 has a spectral cut-off or break below 100~MeV.
Therefore, the absence of a pulsed signal above 1~GeV, the spatial
coincidence and the similar extension of the LAT source with the
pulsar wind nebula as seen in X-rays and very high energy
$\gamma$-rays, strongly suggest that the unpulsed $\gamma$-ray
emission detected by the LAT above 1 GeV is dominated by the PWN.

\begin{figure*}
\begin{center}
\includegraphics[width=\linewidth]{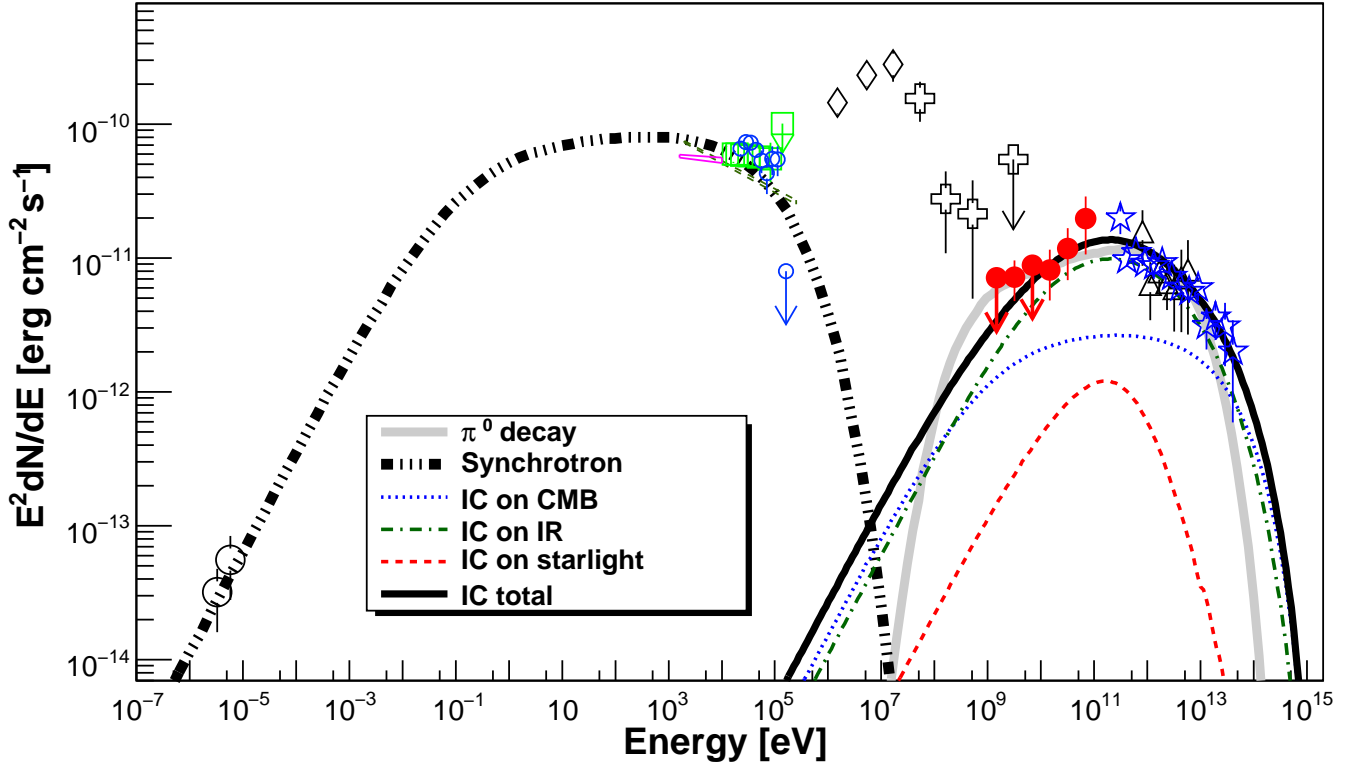}
\caption[]{Spectral energy distribution of the pulsar wind nebula
  powered by PSR~B1509$-$58, from radio to very high energy
  $\gamma$-rays. Predicted spectra as discussed in
  Section~\ref{discussion:neb} are overlaid. The total IC spectrum is
  shown with a solid line while thinner lines denote the individual IC
  components: CMB (dotted), infrared (dot-dashed) and optical
  (dashed). The dot-dot-dot-dashed line indicates the corresponding
  synchrotron emission. A hadronic $\gamma$-ray model is also overlaid
  by a thick gray curve.  Observational data points: ATCA (open
  circles, \cite{gae99}, \cite{gae02}), {\it BeppoSAX}/MECS (magenta thin line
  region, \cite{mineo01}), {\it BeppoSAX}/PDS (light green open squares,
  \cite{for06} and the upper limit is $1~\sigma$), {\it INTEGRAL}/IBIS
  (dark blue open circles, \cite{for06} and the upper limit is $1~\sigma$),
       {\it RXTE}/PCA+HEXTE (dark green dashed line region, \cite{marsden}),
       COMPTEL (open diamonds, \cite{kui99}), EGRET (open crosses,
       \cite{kui99}), {\it Fermi}-LAT (this paper, filled circles),
       H.E.S.S. (open stars, \cite{aha05}), CANGAROO-III (open
       triangles, \cite{nak08}).  Note that COMPTEL and EGRET points
       present the DC emission from this region, which is dominated by
       the central pulsar for this energy range and could be
       contaminated by the nearby $\gamma$-ray pulsar PSR~J1509$-$5850
       recently identified by the LAT~\citep{abd10a}.  }
\label{fig:sed}
\end{center}
\end{figure*}

There are two possible interpretations for the origins of $\gamma$-ray
photons from pulsar wind nebulae, i.e hadronic (from proton-proton
interactions) or leptonic (via the inverse Compton process).  The
multi-wavelength picture of MSH~15$-$52 is presented in
Figure~\ref{fig:sed} using all available data on MSH 15$-$52 as
reported in~\cite{nak08}.  The {\it Fermi}-LAT spectral points,
obtained from the analysis described in Section~\ref{nebula}, provide
new constraints on the model parameters.  A simple one-zone model
described in \cite{nak08} can be used to reproduce the
multi-wavelength spectrum of the PWN.  We use the publicly
distributed \footnote{GALPROP model of cosmic-ray transport:
  http://galprop.stanford.edu/web\_galprop/galprop\_home.html}
interstellar radiation field (ISRF) as described in \citet{por05} as
target photons for inverse Compton scattering.  The ISRF spectra are
modeled and given as a function of cylindrical coordinates in the
Galaxy, for IR photons from interstellar dust grains, optical light
from normal stars, and CMB.  We do not consider the production of
$\gamma$-rays via bremsstrahlung because of the low density of this
region (up to $\sim 0.4$~cm$^{-3}$) as reported in \citet{dub02}.  For
simplicity, escape, energy and adiabatic losses as well as the time evolution
of the magnetic field strength in the PWN are neglected, since the
characteristic age of the pulsar is quite young.  As reported
by~\cite{nak08}, a single power-law electron spectrum does not
reproduce the SED; hence the accumulated electron spectrum used here follows a
broken power-law with an exponential cut-off:

\begin{equation}
\displaystyle \frac{dN_{\rm e}}{dE} \propto \frac{(E/E_{\rm
    br})^{-p_1}}{1+(E/E_{\rm
    br})^{p_2-p_1}}\exp{\Bigl(-\frac{E}{E_{\rm max}}\Bigr)},
\end{equation}
where $E_{\rm max}$, $E_{\rm br}$, $p_1$ and $p_2$ are the maximal
energy, break energy and the indices of the electron spectrum
respectively.  The best fit yields the parameters listed in
Table~\ref{tab:par} and is overlaid in Figure~\ref{fig:sed}.

\begin{table}[h!]
\begin{center}
\begin{tabular}{cc}
\hline
\hline
parameter & value \\
\hline
$p_1$.............. & 1.5 \\
$p_2$.............. & 2.9 \\
$E_{\rm br}$(eV)...... &   $4.6\times 10^{11}$\\
$E_{\rm max}$ (eV).. & $2.6\times 10^{14}$ \\
$B$ ($\mu$G)........ & 17\\
$W_e$ (ergs)..... & $3.0\times 10^{48}$\\
$W_p$ (ergs)..... & $1.2\times 10^{51}(1.0/n)$\\
\hline
$L_{\rm sy}$ (erg/s)... & $3.9\times 10^{36}$\\
$L_{\rm ic}$ (erg/s)... & $3.1\times 10^{35}$\\
\hline
\end{tabular}
\caption{Parameters derived from the multi-wavelength spectral modeling}\label{tab:par}
\end{center}
\end{table}

The fitted mean magnetic field strength of 17~$\mu$G is identical 
with the value suggested by TeV observations \citep{aha05, nak08} 
and consistent with the lower limit of $\ge$ 8~$\mu$G obtained from X-ray data \citep{gae02}. 
We confirm that the $\gamma$-ray emission from the PWN is dominated by
the inverse Compton scattering of the IR photons from interstellar
dust grains with a radiation density fixed at 1.4~eV~cm$^{-3}$ , which is the nominal
value of the GALPROP ISRF. 
Here, the contribution of the optical photon field is negligible
because of the Klein-Nishina effect \citep{kle29}.

The nature of the break energy $E_{\rm br}$ remains unclear.  A
radiation cooling break which is dominated by the synchrotron loss in
this leptonic scenario is a primary candidate.  With the assumed
magnetic field ($B=17~\mu$G) and system age ($\tau=1.7$~kyr), the
break energy can be calculated using standard formulae in \citet{pac70}
as $6\pi m_{\rm e}^2c^3/B^2\sigma _{\rm
  T}\tau \sim$ 24~TeV leading to a break in the photon spectrum at
$36(B/1~\mu {\rm G})(\tau/10^3~{\rm yrs})^{-2}~{\rm eV}\sim$ 0.2~keV.
Even considering the limit of the one-zone
approximation and potential uncertainty in the parameters obtained,
the fitted value of 460~GeV, which predicts a break in the photon
spectrum at $\sim 8\times 10^{-3}$~eV, is much lower than the expected
value. 
However, the photon
index change $\Delta \Gamma =0.7$ is compatible with the prediction by
the usual synchrotron cooling $\Delta \Gamma=0.5$.  One should note
that the synchrotron break in this pulsar wind nebula is expected to
occur just below the X-ray band and no hints of a break have yet
been observed. More likely, this break energy could be an intrinsic
characteristic of the electron spectrum injected in the PWN as
suggested by \cite{dej08}, though $p_2 =2$ is expected.  Another
possibility would be that the spectral break of electrons may
correspond to the energy scale where the electron acceleration
mechanism switches, for instance, from magnetic reconnection
\citep{zen01} to the usual first-order Fermi acceleration.  However,
while \citet{zen01} indeed predict an index of unity, the
production of an index harder than 2 is explained with difficulty by
Fermi acceleration.

The total energy PSR~B1509$-$58 can supply to its PWN is strongly
dependent on its initial spin period $P_0$, which is generally
unknown, as $E_{\rm tot}=2\pi ^2I(\frac{1}{P_0^2}-\frac{1}{P^2})$,
where $I$ is the moment of inertia of the neutron star.  $P_0$ can be
analytically calculated for an ideal case, assuming that $k$ and $n$
are constant in the braking equation $\dot{\Omega} = -k\Omega ^n$
\citep{gae06}.  Using the standard parameters of the pulsar
PSR~B1509$-$58 (period of 150 ms, period derivative of $1.5 \times
10^{-12}\, \rm{s}\,\rm{s}^{-1}$ and braking index
$n=2.84$~\citep{liv05}), we obtain $P_0 = $ 16~ms and $E_{\rm tot} =
7.5 \times 10^{49}$~ergs.  Knowing that the fit requires a total injected
energy (i.e. integrated electron energies above 1~GeV) of 
$W_e = 3.0 \times 10^{48}$~ergs, $\sim 4\%$ of $E_{\rm
 tot}$ should be converted into the current kinetic energy of
electrons.

The equipartition magnetic field strength can be estimated as $B_{\rm
  eq} = \sqrt{8\pi W_e/V}$, where $V$ is a volume of the emission
region. Assuming a spherical region of radius $r$, we obtain $B_{\rm
  eq} = 22(10{\rm pc}/r)^{3/2}~$$\mu$G. In the case of the considered
source where $r$~$\sim$ 10~pc, the PWN is particle dominated as
suggested by \citet{che04}.  The measurement of the extension of the
HE $\gamma$-ray emission is of particular interest to better estimate
the boundary of the PWN.

We also consider a $\pi ^0$ decay model, assuming a proton spectrum
described by a power-law with a cut-off to fit the data points, though
there are few theoretical indications supporting an injection of such
hard protons.  We obtain an index of protons of 1.9 with a cut-off
energy of of 60~TeV, which yields the accumulated energy of protons above 1~GeV of
$1.2\times 10^{51} (1.0~{\rm cm}^3)/n$ ergs, where $n$ is a number
density of target nuclei.  This scenario is highly disfavoured from
the energetics: even with a very high density of $\sim 10$~cm$^{-3}$
as mentioned by \cite{dub02} for the northwest limb of MSH 15$-$52,
the energy required would significantly exceed the total energy that
the pulsar can supply to its nebula ($E_{\rm tot} = 7.5 \times
10^{49}$~ergs).

\section{Conclusion}

We report the detection of pulsed high energy $\gamma$-rays from
PSR~B1509$-$58 below 1~GeV and extended emission from its PWN in
MSH~15$-$52 up to 100~GeV using 1 year of survey data with {\it
  Fermi}-LAT.  The LAT light curve of PSR~B1509$-$58 above 30~MeV
presents two peaks. The $\gamma$-ray pulse located at phase
0.33~$\pm$~0.02 is coincident with the main peak observed in X- and
soft $\gamma$-rays \citep{ulm93,kui99}.  The second peak detected at phase 0.96~$\pm$~0.01
may correspond to the marginal detection reported by \citet{kui99} in
the 10 -- 30~MeV energy range covered by EGRET. 
The high-altitude
emission models have problems explaining the peak separation
measured with {\it Fermi} and the radio peak lagging the first
$\gamma$-peak. A confirmation of a possible precursor in the 1351
MHz radio profile \citep{cra01} might change this interpretation.
The 2~$\sigma$ upper limits derived below 1~GeV confirm the spectral break of
PSR~B1509$-$58 in the 10 -- 30~MeV energy range.  More data are needed
to measure the pulsar spectrum in the LAT energy range. Such estimates
can help constrain the shape of the spectrum from X- to $\gamma$-rays
and disentangle between the emission models in the magnetosphere of
PSR~B1509$-$58.  
Both the high-altitude models \citep{Romani 1996,ven09} and the polar cap
model \citep{harding} can accommodate the severe spectral break and the low {\it Fermi}
limits. Details such as the break energy and/or X-ray spectral shape
are not exactly met.
The extended $\gamma$-ray emission observed by the
LAT above 1~GeV is spatially coincident with the PWN powered by
PSR~B1509$-$58. Its morphology is well modeled by a uniform disk or a
Gaussian distribution. The LAT spectrum of the PWN above 1 GeV is well
described by a power-law with a photon index of (1.57 $\pm$
0.17 $\pm$ 0.13) and a flux above 1 GeV of (2.91 $\pm$ 0.79 $\pm$
1.35) $\times 10^{-9}$~cm$^2$~s$^{-1}$.  LAT analyses of the PWN in
MSH~15$-$52 bring new elements to the discussion on the emission
models responsible for the high to very high emission from this
source. The hadronic $\gamma$-ray scenario is highly disfavoured by
the new LAT observations, as suggested by previous TeV observations
\citep{aha05,nak08}. The multi-wavelength spectrum can be explained by
synchrotron and inverse Compton processes, assuming a broken power-law
spectrum for the electrons.  The spectral break, constrained by
multi-wavelength observations, is likely due to an intrinsic break of
electrons injected from the pulsar wind.  About 4~\% of the pulsar's
loss of rotational energy would be required to power the $\gamma$-rays
detected by the LAT, well in the range observed for other pulsar wind
nebulae.  More data are required to estimate the detailed morphology
of the high energy $\gamma$-ray emission and
better constrain the spectral break in the inverse Compton component.\\

The \emph{Fermi} LAT Collaboration acknowledges generous ongoing support
from a number of agencies and institutes that have supported both the
development and the operation of the LAT as well as scientific data analysis.
These include the National Aeronautics and Space Administration and the
Department of Energy in the United States, the Commissariat \`a l'Energie Atomique
and the Centre National de la Recherche Scientifique / Institut National de Physique
Nucl\'eaire et de Physique des Particules in France, the Agenzia Spaziale Italiana
and the Istituto Nazionale di Fisica Nucleare in Italy, the Ministry of Education,
Culture, Sports, Science and Technology (MEXT), High Energy Accelerator Research
Organization (KEK) and Japan Aerospace Exploration Agency (JAXA) in Japan, and
the K.~A.~Wallenberg Foundation, the Swedish Research Council and the
Swedish National Space Board in Sweden.

Additional support for science analysis during the operations phase is gratefully
acknowledged from the Istituto Nazionale di Astrofisica in Italy and the Centre National d'\'Etudes Spatiales in France.

The Parkes radio telescope is part of the Australia Telescope which is funded by the Commonwealth Government for operation as a National Facility managed by CSIRO. We thank our colleagues for their assistance with the radio timing observations.


\end{document}